\documentstyle[12pt]{article}    
\newcommand{\eq}{\begin{eqnarray}}
\newcommand{\en}{\end{eqnarray}}
\begin{document}
\thispagestyle{empty}

\hfill {\bf BUTP-99/30}

\vspace*{.3cm}
\begin{center}

{\Large\bf Report of Working Group on Electromagnetic Corrections
\footnote{Talk given at Eighth International Symposium on Meson-Nucleon
Physics and the Structure of the Nucleon (MENU99), Zuoz, Engadine, Switzerland,
15 - 21 August 1999.}}

\vspace*{.5cm}

A. Rusetsky\\
{\em Institute for Theoretical Physics, University of Bern,
Sidlerstrasse 5, CH-3012, Bern, Switzerland,\\ 
Bogoliubov Laboratory of Theoretical Physics, Joint Institute
for Nuclear Research, 141980 Dubna, Russia and\\ 
HEPI, Tbilisi State University, 380086 Tbilisi, Georgia}

\end{center}

\vspace*{.4cm}

\begin{abstract}

\noindent
The talks delivered by M. Knecht, H. Neufeld,
V.E. Lyubovitskij, A. Rusetsky and J.~Soto during the session
of the working group of electromagnetic corrections to hadronic processes
at the Eight International Symposium MENU99, cover a wide range of
problems. In particular, those include: construction of the effective
Lagrangians that then are used for the evaluation of electromagnetic
corrections to the decays of $K$ mesons; evaluation of some of the 
low-energy constants in these Lagrangians, using sum rules and the 
large-$N_c$ arguments; complete calculations of electromagnetic corrections 
to the $\pi\pi$ scattering amplitude at $O(e^2p^2)$; 
the general theory of electromagnetic bound states of hadrons 
in the Standard Model.

\end{abstract}

\newpage

The problem of unique disentangling of strong and electromagnetic 
interactions in had\-ro\-nic 
transitions has been a long-standing challenge for 
theorists. All data obtained in high-energy physics experiments, contain a
highly nontrivial interplay of strong and electromagnetic effects, with a
huge difference in the interaction ranges. In addition, there are the
isospin-breaking effects caused by the difference in quark masses that
generally has a non-electromagnetic origin. In the analysis of the 
experimental data, however, one would prefer to unambiguously subtract
all isospin-breaking corrections from the hadronic characteristics 
in order to obtain the quantities defined in "pure QCD", at equal quark
masses -- the case where the most of the theoretical predictions are done.

In the context of $\pi N$ scattering problem, the issue of the 
electromagnetic corrections has been extensively studied by using the
dispersion relations~\cite{Tromborg}, and the potential model~\cite{Rasche}.
The closely related problem of electromagnetic corrections to the 
energy-level shift and decay width to the pionic hydrogen and pionic 
deuterium was analyzed again by using the potential model~\cite{Sigg}.
Both above approaches are based upon the certain assumptions about the
precise mechanism of incorporation of electromagnetic effects in the strong
sector, provided that the details of strong interactions are 
unknown. It remains, however, unclear, how 
to estimate the systematic error caused by these assumptions or, in other
words, how to consistently include within these approaches all 
isospin-breaking effects which are present in the Standard Model.
Given the fact, that the above approaches are used for the partial-wave
analysis of $\pi N$ scattering data and the measurements of $\pi^-p$,
$\pi^-d$ atom characteristics that result in the independent determinations 
of values of the $S$-wave $\pi N$ scattering lengths, the consistent
treatment of the isospin-breaking effects might be helpful for understanding
the discrepancies between the results of different analyses. 

Further, the problem of the systematic treatment of electromagnetic corrections
is crucial for the analysis of experimental data on the decays of $K$ mesons.
In particular, the study of $K_{l4}$ decays that at present time is performed 
by E865 collaboration at BNL~\cite{Pislak}, and by KLOE at DA$\Phi$NE facility
(LNF-INFN)~\cite{deSimone}, will enable one to measure the parameters of the
low-energy $\pi\pi$ interaction and thus provide an extremely valuable
information about the nature of strong interactions at low energy. Preliminary
results~\cite{Pislak} are so far obtained in the negligence of the 
electromagnetic effects which might significantly affect the amplitudes in 
the threshold region.    

Last but not least, a complete inclusion of electromagnetic corrections is
needed in order to fully exploit the high-precision data on 
hadronic atoms provided by the DIRAC collaboration at CERN ($\pi^+\pi^-$), 
DEAR collaboration at DA$\Phi$NE ($K^-p$, $K^-d$) , by the experiments at 
PSI ($\pi^-p$, $\pi^-d$), KEK ($K^-p$), etc. These experiments, that allow
for the direct determination of the hadronic scattering lengths from the
measured characteristics of hadronic atoms: level energies and decay widths,
contribute significantly to our knowledge of the properties of QCD in the
low-energy regime. In particular, the measurement of the difference of the
$S$-wave $\pi\pi$ scattering lengths $a_0-a_2$ by the DIRAC experiment will
allow one to distinguish between the large/small condensate scenarios of
chiral symmetry breaking in QCD: should it turn out that the measured value of
$a_0-a_2$ differs from the prediction of standard ChPT~\cite{ChPT}, one has to
conclude that the symmetry breaking in QCD proceeds differently~\cite{Stern}
from the standard picture. Further,
the $\pi N$ scattering length $a_{0+}^-$ that is measured by the experiments
on pionic hydrogen and pionic deuterium, can be  used as an input
to determine the $\pi NN$ coupling constant, and from the precise knowledge
of $KN$ scattering lengths one might extract the new information
about e.g. the kaon-sigma term and the strangeness content of the 
nucleon.

According to the modern point of view, the low-energy interactions of
Goldstone bosons (pions, kaons...) in QCD can be consistently described
by using the language of effective chiral Lagrangians. The amplitudes of the
processes involving these particles below $\sim 1~{\rm GeV}$  can be 
systematically expanded in series over the external momenta of Goldstone 
particles and light quark masses. All nonperturbative QCD dynamics is than
contained in the so-called low-energy constants of the effective Lagrangian,
and only a finite number of those contribute at a given order in this
expansion. These constants should be, in principle, calculable from QCD. 
However, at the present stage they are considered to be
free parameters to be determined from the fit to the experimental
data~\cite{ChPT,Stern,ChPTs}. The approach can be generalized to the sector 
with baryon number equal to $1$ or $2$~\cite{1N,2N}. 
Moreover, the approach allows
for the systematic inclusion of electromagnetic  interactions -- 
albeit at the cost of the increased number of low-energy constants in the
effective Lagrangian~\cite{Urech,Meissner-em,Rupertsberger}. These new 
"electromagnetic" low-energy constants describe the high-energy
processes corresponding to the direct interaction of quarks with photons,
and thus contribute the missing piece to the potential-type models --
in the latter, generally, only the long-range part of the electromagnetic
interactions, corresponding to the photon exchange between different hadrons,
is taken into account. From the size of the effect coming from the 
"electromagnetic" low-energy constants, along with other missing sources
of electromagnetic corrections (see below), one can therefore have a judgment
on the systematic error caused by the choice of the potential-type models.

From the above, it seems that the most nontrivial task that
one encounters in the systematic treatment of the electromagnetic interactions
in the low-energy hadronic processes, is related to the determination of the
precise values of the "electromagnetic" low-energy constants in the
effective low-energy Lagrangians that, at the present time, are rather
poorly known. Different methods have been used to this end so far.
The resonance saturation method was used~\cite{Baur} to evaluate
these constants in the $O(e^2p^2)$ Lagrangian with Goldsone bosons.
Further, in Ref.~\cite{Moussallam} it was demonstrated that these constants 
can be expressed as a
convolution of a QCD correlation function with the photon propagator,
plus a contribution from the QED counterterms. The sum rules were
then used to evaluate these constants~\cite{Moussallam}. Somewhat different
approach to the calculation of these constants was used in~\cite{Prades}. 
The results of these approaches do not all agree. One important lesson,
however, can be immediately drawn: since these low-energy constants turn
out to be dependent on the QCD scale $\mu_0$ that should be
introduced in the QCD Lagrangian after taking into account the electromagnetic
corrections~\cite{Moussallam,Prades}, the naive separation of the
isospin-breaking effects into the "electromagnetic" and "strong" parts,
the latter corresponding to the difference of quark masses, does not hold in
general and can be carried out only in certain observables at a certain chiral
order. Generally, both these parts are dependent on QCD scale $\mu_0$ that
then cancels in the sum. For this reason, hereafter, we would prefer to
speak about the isospin-breaking corrections to the physical observables,
rather then consider individual contributions to it.  

The session of the working group of electromagnetic corrections at MENU99
symposium was designed to cover all important steps of the treatment of
isospin-breaking corrections on the basis of modern effective theories:
starting from the construction of the effective Lagrangians containing non-QCD
degrees of freedom -- photons and leptons, and from the examples of the
determination of some of the non-QCD low-energy constants, using sum rules
and large $N_c$ arguments, to the actual application of the framework to the
calculation of physical quantities: $P_{l2}$ and $P\rightarrow\ell^+\ell^-$ 
decay rates, $\pi\pi$
scattering amplitudes, as well as observables of $\pi^+\pi^-$ and $\pi^-p$
hadronic atoms -- energy levels and decay characteristics.

A construction of a low-energy effective field theory which allows the
full treatment of isospin-breaking effects in semileptonic weak
interactions, was discussed in the talk of H.~Neufeld~\cite{Neufeld} (see
also~\cite{Rupertsberger} for more details). 
In addition to the pseudoscalars and the photon, 
also the light leptons were included as dynamical degrees of
freedom in an appropriate chiral Lagrangian: only
within such a framework, one will have full control over all
possible isospin breaking effects in the analysis of new high statistics
$K_{\ell 4}$ experiments  by the E865 and KLOE collaborations.
The same methods are also necessary for the
interpretation of forthcoming high precision experiments on
other semileptonic decays like $K_{\ell 3}$, etc.

The one-loop functional in the presence of leptonic sources was evaluated
by using the superheat kernel technique~\cite{Berezinian}.
At next-to-leading order, the list of low-energy constants has to be
enlarged as compared to the case of QCD+photons. If only the terms
at most quadratic in lepton fields, and at most linear in Fermi coupling
$G_F$ are considered, the number of additional low-energy constants
$X_i,~i=1\cdots 7$ is equal to $7$. Further, regarding "pure" lepton
or photon bilinears as "trivial", only three from the remaining five
low-energy constants will contribute to realistic physical processes.   
One may therefore conclude that the inclusion of virtual
leptons in chiral perturbation theory proceeds at a rather moderate cost.

As an immediate application of the formalism, a full set of electromagnetic 
corrections to $P_{l2}$ decays were calculated. It was demonstrated that,
in some specific combinations of widths of different decay processes, the "new"
low-energy constants cancel, thus leading to the predictions
at the one-loop order that do not depend on the parameters $X_i$. 
A full calculation of the electromagnetic corrections
to the $K_{l3},~K_{l4}$ processes is in progress.

As it is evident, the question of actual evaluation
of a large number of low-energy constants lies at a heart of the 
successful application of the effective Lagrangian approach.
In the talk by M.~Knecht, the evaluation of one of such low-energy
constants that contributes to the decays of pseudoscalars into lepton pairs,
was given on the basis of sum rules and large $N_c$ arguments (for more
details, see~\cite{Peris}). In the large $N_c$ limit, the QCD spectrum
consists of a tower of infinitely narrow resonances in each channel.
Employing further the so-called Lowest Meson Dominance (LDM) approximation that
implies the truncation of the infinite sum over resonances in the sum rules,
one can relate the low-energy constants to the known parameters of the
low-lying resonances. It was demonstrated that using the value of the
particular low-energy constant determined in the LDM approximation leads to
the values of ratio of branching ratios
$Br(P\rightarrow\ell^+\ell^-)/Br(P\rightarrow\gamma\gamma)$
for the processes $\pi^0\rightarrow e^+e^-$ and 
$\eta\rightarrow \mu^+\mu^-$ that are 
consistent with the present experimental data.
The predictions for the same value in the process $\eta\rightarrow e^+e^-$
were also given.

As another application of the effective theories, in the talk by 
M.~Knecht~\cite{Knecht} (see also~\cite{KU}) a complete calculation of the
electromagnetic corrections to the $\pi^0\pi^0\rightarrow\pi^0\pi^0$
and $\pi^+\pi^-\rightarrow\pi^0\pi^0$ scattering amplitudes at
$O(e^2p^2)$ has been performed. The latter case is particularly interesting
since the the results can be directly translated into the corrections to the
decay width of the $\pi^+\pi^-$ atom. It turns out that the size of
isospin-breaking corrections to the $\pi\pi$ scattering amplitudes is
of the same order of magnitude that the two-loop strong corrections and,
therefore, can not be neglected.

Bound states of hadrons in chiral effective theories -- hadronic atoms --
were considered in the talks by J. Soto, V. Lyubovitskij and
A. Rusetsky~\cite{Soto,Lyubovitskij,Rusetsky}. It was demonstrated that,
a systematic evaluation of isospin-breaking corrections to the observable
characteristics of this sort of bound systems is possible order by order
in ChPT. To this end, a nonrelativistic effective Lagrangian approach
was used, that provides the necessary bridge between the bound-state
characteristics and the scattering $S$-matrix elements in a most elegant and
economical manner: at the end, according
to the matching condition of relativistic and nonrelativistic theories,
these characteristics are expressed in terms of the scattering matrix 
elements calculated in the relativistic theory, and to the latter one
can apply the conventional machinery of ChPT.

The talk by J. Soto~\cite{Soto} (details -- in Ref.~\cite{Eiras}) 
was focused on the foundations of
the nonrelativistic effective Lagrangian approach as applied to the
$\pi^+\pi^-$ atom decay problem. Different scales relevant for the problem
of interest have been thoroughly discussed and disentangled. It has
been demonstrated that, matching the parameters of the nonrelativistic
Lagrangian to the relativistic theory, it is possible to evaluate the
decay width of the $\pi^+\pi^-$ atom order by order in ChPT. 

The nonrelativistic effective Lagrangian approach was applied to the 
$\pi^+\pi^-$ atom decay problem as well in the talk by 
V. Lyubovitskij~\cite{Lyubovitskij} (see~\cite{Bern} for more details).
It was demonstrated, however, that employing the different technique for
matching of relativistic and nonrelativistic theories, it is possible
to obtain the general expression for the decay width of the $\pi^+\pi^-$
atom in the first nonleading order in isospin breaking, without an explicit
use of chiral expansion -- that is, the result is valid in all orders
in ChPT. At order $O(e^2p^2)$ in ChPT, one may use the results of 
calculations by Knecht and Urech~\cite{KU} -- in this manner, it was
demonstrated that the one-loop corrections to the decay width are rather
small, including the uncertainty coming from the "strong" and 
"electromagnetic" low-energy
constants. The bulk of total correction is given already by the tree-level
diagram that is free from this uncertainty. The results given in the talk 
by V. Lyubovitskij, finalize the treatment of the $\pi^+\pi^-$ atom decay
problem: the width is now known to a sufficient accuracy that allows one to
fully exploit the future precision data from DIRAC experiment at CERN.

In the talk by A. Rusetsky~\cite{Rusetsky} the nonrelativistic effective
Lagrangian approach was applied to the calculation of the energy of
the ground state of the $\pi^-p$ atom. At this example, one can fully
acknowledge the might and flexibility of the nonrelativistic approach: the
spin-dependent part of the problem trivializes, and the treatment proceeds
very similarly to the case of the $\pi^+\pi^-$ atom. However, the 
isospin-breaking piece of the relativistic scattering amplitude in the 
$\pi^-p$ case already at the tree level (order $p^2$) contains both
"strong" and "electromagnetic" low-energy constants. This part of the
isospin-symmetry breaking effect is missing in the potential 
model~\cite{Sigg}. From the explicit expression of the $\pi^-p$ scattering
amplitude at $O(p^2)$, one can immediately identify two distinct sources
of isospin-breaking corrections that are not present in the potential model.
The terms that depend on the quark masses in the strong part of the $\pi^-p$
amplitude, contribute to the isospin-breaking piece -- this contribution
is proportional to charged and neutral pion mass difference. In addition, the
direct quark-photon interaction that is encoded in the "electromagnetic"
low-energy constants, also contributes to the isospin breaking. Uncertainty
introduced by the poor knowledge of these low-energy constants is, unlike the
$\pi^+\pi^-$ case, much larger as compared to the estimate based on the
potential model -- thus, the latter considerably underestimates the 
systematic error in the analysis of the pionic hydrogen data provided by the
experiment at PSI.

To summarize, I shall briefly dwell on the developments that are foreseen
in the nearest future.

$\bullet$ A complete treatment of the isospin-breaking corrections to
$K_{l3}$ and $K_{l4}$ decays on the basis of modern effective field theories
that are now in progress, is of a great importance for the analysis of
precision data samples from E865 (BNL) and KLOE (LNF-INFN) experiments.

$\bullet$ For the systematic treatment of the isospin-breaking corrections,
the knowledge of the precise values of the low-energy constants that enter 
the effective Lagrangian, is necessary. In particular, this concerns the
values of "electromagnetic" low-energy constants which are poorly known.
Activities based on sum rules,
resonance saturation models, etc. may provide an extremely useful 
information in this respect. 

$\bullet$ At present, the problem of $\pi^+\pi^-$ atom decay is completely
understood, both conceptually and numerically. The expected high-precision
data from DIRAC experiment, therefore, can be used to determine the
difference of the $\pi\pi$ $S$-wave scattering lengths $a_0-a_2$ quite
accurately. In the contrary, the issue of the isospin-breaking corrections
to the observables of pionic hydrogen (and pionic deuterium) needs to be
further investigated. The uncertainty due to the poor knowledge of
low-energy constants in the isospin-breaking part of the $\pi N$ scattering 
amplitude is large already at tree level, and the evaluation of one-loop
contributions are in progress. These studies will be even more important
for the future improved experiment at PSI~\cite{Gotta} that intends to
measure the $S$-wave $\pi N$ scattering lengths, using the data from the
pionic hydrogen alone. A detailed investigation of the properties of the
kaonic atoms which will be studied by the DEAR collaboration at DA$\Phi$NE,
is planned.

$\bullet$ The success of the nonrelativistic Lagrangian approach to the
hadronic atom problem clearly demonstrates that the formalism of
the potential model based on Schr\"{o}dinger-type equations, can be 
applied to the evaluation of the
isospin-breaking corrections, provided the potential contains a full content
of isospin-symmetry breaking effects of the Standard Model. Therefore,
it will be extremely important to set up a systematic framework for the
derivation of
the potentials from the effective field-theoretical Lagrangians -- then,
the already existing machinery of the potential model can be directly 
exploited. 

\vspace*{.3cm}

{\it Acknowledgments}. 
I would like to thank the organizers of MENU99 symposium for their large
effort, that created an atmosphere of intense discussions and learning, and
J. Gasser for reading the manuscript.
This work was supported in part by the Swiss National Science
Foundation, and by TMR, BBW-Contract No. 97.0131  and  EC-Contract
No. ERBFMRX-CT980169 (EURODA$\Phi$NE).

\bibliographystyle{unsrt}

\end{document}